\documentclass{article}
\usepackage{xcolor}
\newcommand{\lovric}[1]{{\color{blue} $\triangleright$ #1}}
\newcommand{\vekk}[1]{}

\begin{document}

\begin{center}

{\large \bf PHILOSOPHICAL FOUNDATIONS OF STATISTICS}\\[4mm]

{\large \it Inge S. Helland \footnote{Inge Svein Helland is retired professor from University of Oslo. He was Head of Department of Mathematical Sciences, Agricultural University of Norway (1987--1989) and Head of Statistics Division, Department of Mathematics, University of Oslo (1999--2001) and professor at the same division until retirement. He was Associate Editor, \textit{Scandinavian Journal of Statistics} (1994--2001). Professor Helland has (co-)authored about 75 papers, reports, manuscripts, including the books \textit{Steps Towards a Unified Basis for Scientific Models and Methods} (World Scientific, Singapore, 2010) and \textit{Epistemic Processes. A Basis for Statistics and Quantum Theory. Second Edition} (Springer, 2021)}} \\[1mm]

Professor emeritus, Department of Mathematics, University of Oslo, Norway


{\large {\it {Nils Lid Hjort \footnote{Nils Lid Hjort 
is professor at Department of Mathematics, University of Oslo.
He has co-authored {\it Model Selection} (2008), 
{\it Bayesian Nonparametrics} (2010), 
{\it Confidence, Likelihood, Probability} (2016) with Cambridge
Unviversity Press, and {\it Highly Structured Stochastic Systems} (2003)
with Oxford University Press. He has published widely on 
methodological and applied themes, including model selection
and averaging, density and regression estimation, 
survival and event history modelling, Bayesian and frequentist
nonparametrics, robust parametrics, empirical likelihood, 
goodness-of-fit, changepoints and trends identification, 
whale ecology, and peace and conflict studies. }}}}\\[1mm]

Professor, Department of Mathematics, University of Oslo, Norway

{\large {\it {Gunnar Taraldsen \footnote{Gunnar Taraldsen is professor at Department of Mathematical Sciences, NTNU. His current research interest focusses on the foundational principles of statistical inference including quantum probability, artificial intelligence, and machine learning. The aim is to bridge foundations for statistical inferences, to facilitate objective and replicable scientific learning, and to develop analytic and computing methodologies for data analysis.
Earlier research includes quantum mechanics, mathematical physics, measure theory, probability, medical ultrasound, non-linear wave propagation, numerical acoustics, medical statistics, outdoors sound propagation, community noise annoyance, acoustic noise mapping, underwater localization, underwater wave propagation, and spherical microphone arrays.}}}}\\[1mm]

Professor, Department of Mathematical Sciences, Norwegian University of Science and Technology

\end{center}

The philosophical foundations of statistics involve issues in theoretical statistics, such as goals and methods to meet 
these goals, and interpretation of the meaning of inference using statistics. They are related to the philosophy of science 
and to the \lovric{philosophy of probability}.

As with any other science, the philosophical foundations 
of statistics are closely connected to its history (\lovric{Statistics, History of}), which again is connected to the men 
who shaped it. Some of the most important names in this connection are Thomas Bayes (1701--1761), Sir Ronald A. Fisher (1890--1962),
Sir Harold Jeffreys (1891–-1989),
and Jerzy Neyman (1894--1981). 

The standard statistical paradigm is tied to the concept of a statistical model
defined by a family of probability measures  $P^{\theta}(\cdot)$
on the observations indexed by the model parameter $\theta$.
Inference is done on the model parameter space.
This paradigm was challenged by Breiman (2001), who argued for an 
algorithmical, more intuitive model concept. According to Breiman, the statistical community has been committed to the almost exclusive use of data models. This commitment has led to irrelevant theory, questionable conclusions, and has kept statisticians from working on a large range of interesting current problems. Algorithmic modelling, both in theory and practice, has developed rapidly in fields outside statistics. So he concludes that: If our goal as a field is to use data to solve problems, then we need to move away from exclusive dependence on data models and adopt a more diverse set of tools. In modern times, the interface between statistics and machine learning, see below, has moved us closer to Breiman's goal.

Breiman's tree models are still much in use, together with 
other algorithmical bases for inference, for instance within chemometry. For an attempt to explain some of these 
based on the framework of the ordinary statistical paradigm, see Helland (1990). The relevant algorithm, partial least squares regression (\lovric{Partial Least Squares Regression Versus Other Methods}), and the resulting model, has now been generalized to what is called envelope models, see Cook (2018).

It is an important observation that not all indexed families 
of distributions lead to sensible models. McCullagh (2002) 
showed that mathematically well-defined models may lead
to different degrees of absurdity. Attempting to formulate
principles of `common sense' for statistical modelling, 
to fully safeguard against all potential absurdities, 
remains an elusive enterprise, however. 

Historically, the first statistical methods were developed 
under the assumption that data were normally distributed. 
This assumption, if not exactly true, can sometimes be motivated
by considerations of asymptotics and approximations. 
The assumption can also be tested (\lovric{Normality Tests}). 
Linear models, including regression (\lovric{Linear Regression Models}) 
and \lovric{analysis of variance} models, are particularly important. 
The methodology and applicability of exponential families 
of distributions have been greatly extended and advanced 
over the past few decades, see e.g.~Efron (2022).
(\lovric{Exponential Family Models}) 
In addition to normal distributions, this class of distributions 
includes  Poisson, binomial, gamma, negative binomial and 
inverse Gaussian distributions. Modern developments 
include those for \lovric{generalized linear models}, 
for instance \lovric{logistic regression}. 

The standard statistical model concept can be extended by implementing some kind of model reduction principles (Wittgenstein, 1921: `The process of induction is the process of assuming the simplest law that can be made to harmonize with our experience') or by, e.g., adjoining a symmetry group to the model (Helland 2004, 2009).

To arrive at methods of inference, the model concept must be supplemented by certain principles. In this 
connection, an experiment is ordinarily seen as given by a statistical model together with some focus parameter.
Many statisticians agree at least to some variants of the following three principles: The conditionality principle (When you choose an 
experiment randomly, the information in this large experiment, including the randomization, is not more than the information in the 
selected experiment), the sufficiency principle (Experiments with equal values of a \lovric{sufficient statistic}, have equal 
information) and the likelihood principle (All the information about the parameter is contained in the \lovric{likelihood} for 
the parameter, given the observations). Birnbaum's famous theorem says that the likelihood principle follows from the  
conditionality principle together with the sufficiency principle (for some precisely defined version of these principles). 
This, and the principles themselves, are discussed in detail in Berger and Wolpert (1984).

Berger and Wolpert (1984) also argue that the likelihood principle `nearly' leads to Bayesian inference, as the only mode of inference 
which really satisfies the likelihood principle. This whole chain of reasoning has been countered by Kardaun et al. (2003) 
and by leCam (1984), who states that he prefers to be a little `unprincipled'.

It is argued in Helland (2021) that the likelihood principle is 
less controversial when it is linked to a fixed context. 
This can for instance also be coupled to an example where 
different designs, binomial and geometrical, with the same 
parameter $p$ and the same likelihood, lead to different 
inferences, different confidence intervals. Other setups where details of 
the experimental design enter the best confidence distributions, 
beyond likelihood functions, are discussed in Schweder and Hjort (2016). 

In Helland (2021) a focused likelihood principle is developed, formulated in terms of an operator, which can be taken as part of a basis for developing quantum mechanics. The relationship in general between the foundations of quantum theory and of statistical theory is further discussed there. According to Niels Bohr, concepts are important when communicating, but in basic physics, concepts are used to describe relations, not only facts (data). `Everything we can call real, is made up of things that cannot be regarded as real' is a citation from Bohr. In statistics, this can be taken to mean that everything depends on our mental models.

To arrive at statistical inference, whether it is point estimates (\lovric{Estimation}, \lovric{Estimation, An Overview}), \lovric{confidence intervals} (credibility intervals 
for Bayesians) or hypothesis testing, we need some decision theory (\lovric{Decision Theory: An Introduction}, \lovric{Decision Theory: An Overview}). Such decision theory may be formulated differently 
by different authors. The foundation of statistical inference from a Bayesian point of view is discussed by Good (1988), 
Lindley (2000) and Savage (1972). From the frequentist point of view it is argued by Efron (1986) that one 
should be a little more informal; but note that a decision theory may be very useful also in this setting.

The philosophy of foundations of statistics involves many further questions which have direct impact on the theory 
and practice of statistics: Conditioning, \lovric{randomization}, shrinkage, subjective or objective priors, reference priors, the role of information, 
the interpretation of probabilities, the choice of models, optimality criteria, nonparametric (\lovric{Nonparametric Statistical Inference}) versus parametric inference, the principle of 
an axiomatic foundation of statistics etc. Some papers discussing these issues are Cox (1997) with discussion, Kardaun et al. (2003) and Efron (1978, 1979). 
The last paper takes up the important issue of the relationship between statistical theory and the ongoing revolution of computers.

The struggle between Bayesian statistics and frequentist statistics is is perceived of as less severe today, and as having less practical importance, compared to what can be seen from academic papers before say year 2000. 
The two schools often lead to similar results. As hypothesis testing and confidence intervals are concerned, the 
frequentist school and the Bayesian school must be adjoined by the Fisherian or fiducian school, which has been out of fashion for some time, but has now been revitalized by several authors. 
The question of whether these three schools in some sense can agree on testing, is addressed by Berger (2003).

More generally, one can define three different modes of inference given by Bayesian inference, \lovric{Fiducial inference}, and Frequentist inference.
Bayesian inference is given by computing the unique posterior distribution resulting from the observations, a statistical model for the observations, and a prior distribution.
The result is then an update of the prior state of knowledge into a posterior state of knowledge given the observations.
This mode of inference includes the possibility of improper distributions for both the prior and the posterior as explained by Taraldsen et al. (2022). 

Fiducial inference is given by computing a posterior state of knowledge (the fiducial distribution) based on the observations and a data generating model for the observations.
Different data generating models can give identical statistical models for the observations,
but different fiducial distributions as exemplified by Taraldsen (2021). 
This means that an assumed data generating model contains more information than an assumed statistical model by itself.

Frequentist inference can be seen as the application of a measurement instrument together with
statements regarding the properties of the measurement instrument.
The conclusion of the inference is then not directly a statement about the state of knowledge for some unknown parameter
as for Bayesian and Fiducial inference.
The uncertainty is instead given implicitly by stating properties of the sampling distribution
of the statistic used in the inference (= the measurement instrument).
Fiducial and Bayesian methods can be used to invent methods for frequentist inference,
and in certain cases the methods are optimal (Taraldsen and Lindqvist, 2013).

One purpose of using probability theory in the modelling of experiments is
the possibility of arriving at precise statements regarding uncertainty.
In this context it is most important to know that the
ISO Guide to the Expression of Uncertainty in Measurement gives a standard for this.
This can be compared with the importance of having a standard for how the SI units for length, mass etc are standardized.  
Taraldsen (2006) exemplify that both conventional and Bayesian statistics give a consistent interpretation of the
ISO Guide to the Expression of Uncertainity in Measurement.
The two approaches supplement each other: frequentist statistics estimates the uncertainty of the measurement procedure,
while Bayesian statistics gives the uncertainty of the measurand (= the focus parameter).
More generally, it follows that there is no contradiction between the three modes of inference since
the assumed basis for inference and the interpretations are different.

The field of design of experiments (\lovric{Design of Experiments, A Pattern of Progress}) also has its own philosophical foundation, touching upon practical issues like randomization, 
blocking, and replication, and linked to the philosophy of statistical inference. A good reference here is  Cox and Reid (2000).

In the last decade, statistics has been strongly influenced by and coupled to machine learning, which can be seen as a subfield of artificial intelligence; for a comprehensive introduction of the latter field, see Russell and Norvig (1995). The key common concept can be said to be that of learning, supervised learning, where the goal is to use inputs to predict the values of outputs, and unsupervised learning, which may be called `learning without a teacher'. A good book discussing these issues from a statistical point of view is Hastie et al. (2009). Neural networks can be seen as an important supervised learning procedure in machine learning, while various types of cluster analysis exemplify unsupervised learning.

The philosophical foundations of statistics are concerned with the theoretical and conceptual foundations of statistical methods, and the interpretation of statistical results. Machine learning, on the other hand, is a field of computer science that focuses on the development of algorithms and techniques that enable computers to learn from data and make predictions or decisions. So while machine learning and statistics are closely related and often overlap in practice, they thus have different philosophical foundations.

That being said, the widespread use of machine learning techniques in various fields, including data analysis and modelling, has had an impact on the way that statistical methods can be used and understood. Machine learning has also raised new philosophical questions about the nature of learning and prediction, and how these processes can be understood and studied.

As a general point, it is important to discuss how statistics 
as a science is related to other scientific fields. 
Machine learning, chemometry and quantum mechanics 
have been mentioned above. A common ingredient for all these 
fields is that of mathematical models; see Helland (2022) 
for a general discussion. One scientific area which 
is very closely tied to but not identical to statistics, 
using different models, is that of causality; 
see e.g.~Pearl (2000), Aalen and Frigessi (2007).  
A simple, but fundamental insight is that correlation does not 
imply causation. Causal models are often tied to graphs, 
where arrows are interpreted as assumed causal connections. 
(\lovric{Causation and Causal Inference})
The Sveriges Riksbank Prize in Economic Sciences 
in Memory of Alfred Nobel for 2021 was awarded 
J.D.~Angrist and G.W.~Imbens for 
`for their methodological contributions to the analysis 
of causal relationships'. 

Every statistical analysis depends on a number of decisions, 
made by one or more researchers, often driven by context and 
perhaps convenience. And in every human decision there will 
be some random noise (Kahneman et al., 2021). One may perhaps 
learn something new by modelling this noise. But this belongs 
to the future.

\vspace{1.5cc}

\noindent{\bf References}
\newcounter{ref}
\begin{list}{\small [\,\arabic{ref}\,]}{\usecounter{ref} \leftmargin 4mm \itemsep -1mm}
{\small

\item 
Aalen, O.O. and Frigessi, A. (2007). 
What can Statistics Contribute to a Causal Understanding?
{\it Scandinavian Journal of Statistics}, {\bf 34}, 155--168.

\item 
Berger, J.O. (2003). Could Fisher, Jeffreys and Neyman have agreed on testing? \textit{Statist. Science}, \textbf{18}, 1--32

\item 
Berger, J.O. and Wolpert, R.L. (1984). \textit{The Likelihood Principle.} Lecture Notes, Monograph Series, \textbf{6}, 
Institute of Mathematical Statistics, Hayward.

\item 
Breiman, L. (2001). Statistical modelling: The two cultures. \textit{Statistical Science} \textbf{16}, 199--231.

\item
Cook, R.D. (2018). \textit{An Introduction to Envelopes. Dimension Reduction for Efficient Estimation in Multivariate Statistics.} Wiley, Hoboken, New Jersey.

\item 
Cox, D.R. (1997). The current position of statistics: A personal view. \textit{Intern. Statist. Review,} \textbf{65}, 261--290.

\item 
Cox, D.R. and Reid, N. (2000). \textit{The Theory of Design of Experiments}. Chapman and Hall/CRC, Boca Raton, Fla.

\item 
Efron, B. (1978). Controversies in the foundations of statistics. \textit{Am. Matem. Monthly}, \textbf{85}, 231--246.

\item 
Efron, B. (1979). Computers and the theory of statistics: Thinking the unthinkable. \textit{Siam Review}, \textbf{21}, 460--480.

\item 
Efron, B. (1986). Why isn't everyone Bayesian? \textit{Amer. Statistician}, \textbf{40}, 1--5.

\item
Efron, B. (2022). \textit{Exponential Families in Theory and Practice.} Cambridge University Press, Cambridge.

\item 
Good, I.J. (1988). The interface between statistics and philosophy of science. \textit{Statist. Science}, \textbf{3}, 386--412.

\item
Hastie, T., Tibshirani, R. and Friedman, J. (2009). \textit{The Elements of Statistical Learning. Data Mining, Inference, and Prediction} 2. edition. Springer, Berlin.

\item
Helland, I.S. (1990). Partial least squares regression and statistical models. \textit{Scand. J. Statist.} \textbf{17}, 97-114.

\item 
Helland, I.S. (2004). Statistical inference under symmetry. \textit{Intern. Statist. Review} \textbf{72}, 409--422.

\item 
Helland, I.S. (2009). \textit{Steps Towards a Unified Basis for Scientific Models and Methods.} World Scientific, Singapore. 

\item
Helland, I.S. (2021). \textit{Epistemic Processes. A Basis for Statistics and Quantum Theory.} 2. edition. Springer, Berlin.

\item
Helland, I.S. (2022). On the diversity and similarity of mathematical models in science. \textit{Amer. Rev. Math. Statist.} \textbf{10} (1), 1-10.

\item
Kahneman, D., Sibony, O. and Sunstein, C.R. (2021). \textit{Noise. A Flaw in Human Judgement.} Brockman, Inc., New York.

\item 
Kardaun, O.J.W.F., Salom\'{e}, D., Schaafsma, W., Steerneman, A.G.M., Willems, J.C. and Cox, D.R. (2003). 
Reflections on fourteen cryptic issues concerning the nature of statistical inference. \textit{Int. Statist, Review}, \textbf{71}, 277--318.

\item 
leCam, L. (1984). Discussion in Berger and Wolpert, 182--185.2

\item 
Lindley, D.V. (2000). The philosophy of statistics. The Statistician \textbf{49}, 293--337.

\item 
McCullagh, P. (2002). What is a statistical model? \textit{Ann. Statistics} \textbf{30}, 1225--1310.

\item
Pearl, J. (2000) \textit{Causality. Models, Reasoning, and Inference} 2. edition. Cambridge University Press, Cambridge.

\item
Russell, S. and Norvig, P. (1995). \textit{Artificial Intelligence. A Modern Approach.} Prentice Hall, Upper Saddle River, New Jersey.

\item 
Savage, L.J. (1972). \textit{The Foundations of Statistics}, Dover, New York.

\item 
Schweder, T. and Hjort, N.L. (2016).
{\it Confidence, Likelihood, Probability.}
Cambridge University Press. 

\item 
Taraldsen, G. (2006). Instrument resolution and measurement accuracy. \textit{Metrologia} \textbf{43}, 539-544.

\item 
Taraldsen, G. and Lindqvist, B. (2013). Fiducial theory and optimal inference \textit{The Annals of Statistics} \textbf{41}, 323-341.

\item 
Taraldsen, G., Tufto, J. and Lindqvist, B. (2022). Improper priors and improper posteriors \textit{Scand. J. Statist.} \textbf{49}, 969-991.

\item 
Taraldsen, G. (2021). The Confidence Density for Correlation. \textit{Sankhya A} https://doi.org/10.1007/s13171-021-00267-y. 

\item 
Wittgenstein, L. (1921). \textit{Tractatus Logico-Philosophicus} (tr. Pears and McGuinnes).  Routledge Kegan Paul, London.
}
\end{list}

\end{document}